\documentclass[12pt]{iopart}
\usepackage{iopams}
\usepackage{graphicx}

\begin{document}

 \title{Critical velocity for a toroidal Bose-Einstein condensate flowing through a barrier}

\author{F. Piazza$^1$, L. A. Collins$^2$, A. Smerzi$^3$}
\address{$^1$ Technische Universit\"at M\"unchen, Physik Department, James-Franck-Straße, 85748 Garching, Germany\\
$^2$Theoretical Division, Mail Stop B214, Los Alamos National Laboratory, Los Alamos, New Mexico 87545, USA\\
$^3$INO-CNR and LENS, Largo Enrico Fermi 6, I-50125 Firenze, Italy}


\begin{abstract}
We consider the setup employed in a recent experiment \cite{nist_2011} devoted to the study of the instability of the superfluid flow of a toroidal Bose-Einstein condensate in presence of a repulsive optical barrier. 
Using the Gross-Pitaevskii mean-field equation, we observe, consistently with what we found in \cite{piazza_2009}, that the superflow with one unit of angular momentum becomes unstable at a critical strength of the barrier, and decays through the mechanism of phase slippage performed by pairs of  vortex-antivortex lines annihilating.
While this picture qualitatively agrees with the experimental findings, the measured critical barrier height is not very well reproduced by the Gross-Pitaevskii equation, indicating that thermal fluctuations can play an important role \cite{mathey_2012}.
As an alternative explanation of the discrepancy, we consider the effect of the finite resolution of the imaging system.
At the critical point, the superfluid velocity in the vicinity of the obstacle is always of the order of the sound speed in that region, $v_{\rm barr}=c_{\rm l}$. In particular, in the hydrodynamic regime (not reached in the above experiment)
, the critical point is determined by applying the Landau criterion inside the barrier region. On the other hand, the Feynman critical velocity $v_{\rm f}$ is much lower than the observed critical velocity. We argue that this is a general feature of the Gross-Pitaevskii equation, where we have $v_{\rm f}=\epsilon\ c_{\rm l}$ with $\epsilon$ being a small parameter of the model. Given these observations, the question still remains open about the nature of the superfluid instability.
\end{abstract}

\maketitle

\section{Introduction}

Recent experiments performed at NIST \cite{nist_2007,nist_2011} and at the Cavendish Laboratory \cite{hadzibabic_2012} have demonstrated the existence of persistent currents in a toroidally trapped Bose-Einstein condensate (BEC). Such a setup, apart from showing this hallmark manifestation of superfluidity \cite{leggett_rmp_1999}, has also allowed, with the addition of a repulsive optical barrier, the construction of a closed-loop atom circuit with a weak link \cite{nist_2011}. This circuit constitutes the basic building block for the realization of an ultracold atomic analog of superconducting/superfluid quantum interference devices such as the sensitive magnetic sensors already realized with superconductors (SQUIDs) \cite{clarke_2004} or rotational sensors with superfluid Helium (SHeQUIDs) \cite{packard_2012}. A proof of principle BEC rotation sensor has been very recently realized at NIST \cite{nist_2012}, where the rotation of the reference frame is mimicked by sweeping the barrier around the torus.
Aside from the exciting technological applications, the toroidal setup proves the most suitable for the investigation of the critical velocity of a superfluid moving across an obstacle, since, compared to other setups \cite{ketterle_1999,ketterle_2000,engels_2007}, it has no inhomogeneities along the direction of flow and does not require moving the obstacle given that the condensate can move in the lab frame. The study of superfluid critical velocity and its decay mechanism is a long standing theoretical problem from the early experiments with superfluid Helium \cite{varoquaux_2006} and has yet to be fully understood.

In this manuscript, we model the NIST ultracold atom circuit experiment \cite{nist_2011} using the Gross-Pitaevskii (GP) equation. As already predicted in \cite{piazza_2009}, we find that the superflow with one unit of angular momentum becomes unstable at a critical strength of the barrier and decays through the mechanism of phase slippage through pairs of vortex-antivortex lines parallel to the torus axis.
While this picture qualitatively agrees with the experimental findings, the measured critical barrier height is not well reproduced by the GP equation. This may indicate that thermal fluctuations play an important role by triggering the vortex nucleation before the zero temperature critical velocity predicted by the GP equation is reached. Using a weak link inside a linear waveguide with periodic boundary conditions, the superfluid decay through thermally nucleated vortices is studied in \cite{mathey_2012} by means of the truncated Wigner method. 

As an alternative explanation of the dicrepancy between the GP equation and the experimental data, we consider the effect of the finite resolution of the imaging system.
The comparison indeed depends strongly on the precision with which local density and velocity of the fluid inside the barrier region are experimentally determined and is therefore particularly affected by the finite imaging resolution.
Taking into account that the experimental data already contain a correction due to finite imaging resolution, we estimate the minimum size of a possible further resolution error which could produce agreement.

The ability to describe this setup using the GP equation can prove to be very useful since, as motivated above, the closed loop superfluid circuit is of deep technological and foundational interest. 
Firstly, the GP equation can be employed as an efficient tool to model quantum interference devices, exploiting the versatility of ultracold atomic gases such as the fine tunability of the trapping potential, barrier potential, and the atom-atom interactions. For instance, as demonstrated in \cite{nist_2012}, the ability to tune the barrier height is a means to dynamically change the critical velocity, which constitutes an advantage of the atom circuit over the superconducting/superfluid counterparts.
Secondly, the GP equation allows for a simple theoretical understanding of open questions regarding superfluid instability, which will be the subject of the present manuscript. 

Indeed, after discussing the comparison with the experiment, we then focus on the existence of a general instability criterion determining the critical velocity and on the mechanism underlying the superfluid decay.
At the critical point, we observe that the superfluid velocity in the vicinity of the obstacle is always of the order of the local sound speed, $v_{\rm barr}=c_{\rm l}$, which is a typical feature of the GP equation, at least close to the hydrodynamic regime \cite{rica_1992,adams_2000,watanabe_2009,piazza_2011}. The local sound speed $c_{\rm l}$ is the propagation speed of phonons inside the barrier region, assuming a local density approximation along the flow direction. 

On the other hand, we find that the Feynman critical velocity $v_{\rm f}$ \cite{feynman_1955,varoquaux_2006} is instead much lower than the true critical velocity. The Feynman critical velocity is calculated as the velocity at which the energy of the state with a vortex (in the present case a vortex ring) becomes smaller than the energy of the stationary state without vortices.
Most importantly, we show that in general $v_{\rm f}=\epsilon\ c_{\rm l}$ where $\epsilon$ is a small parameter of the model, that is, the Feynman prediciton can never work in cases where the critical velocity is of the order of the sound speed. 

In particular, in the hydrodynamic regime (not reached in the above experiment) with the fluid locally homogeneous along the flow, we argue that in general the instability criterion corresponds to that of Landau applied in the region where the fluid velocity is the highest, which in the present setup occurs inside the barrier region. The corresponding excitations can thus be the seeds from which vortices grow \cite{anglin_2001}.
At the same time, we have indications that the instability might be of a dynamical nature, since i) we do not have any defect inside the barrier region, and ii) the time scale of vortex nucleation is fast. 
We note that the nature of the phase slip instability, energetic or dynamical, as well as its relation with the Feynman's criterion are not well understood. We conclude by summarizing the current level of understanding and emphasizing the open problems. 

\section{Model}

We model the experiment performed at NIST \cite{nist_2011} using the Gross-Pitaevskii (GP) equation in three spatial dimensions. In dimensionless form, using radial harmonic trap units of length $d_x=2\ \mu{\rm m}$, time $\omega_x^{-1}=1.45\ {\rm ms}$, and energy $\hbar\omega_x=h\times 110\ {\rm Hz}$, the GP equation reads:
\begin{equation}\label{GP}
i \frac{\partial\psi(\boldsymbol{r},t)}{\partial t}=\left[-\frac{1}{2} \nabla^2 +V(\boldsymbol{r},t)+g|\psi|^2\right]\psi(\boldsymbol{r},t)
\end{equation}
where the external potential $V(\boldsymbol{r},t)=V_{\rm tr}(\boldsymbol{r})+V_{\rm barr}(\boldsymbol{r},t)$, with $g=4\pi a_s$ ($a_s$ is the $s$-wave scattering length), and $V_{\rm tr}(x,y,z)=(\omega_z^2/2)z^2+(1/2)(\sqrt{x^2+y^2}-R_0)^2$, where the torus radius $R_0=10$ and the vertical trap frequency $\omega_z=5$. The results presented in the following are obtained by starting with an inital superfluid flow with angular momentum per atom $\ell$ equal to one and $V_{\rm barr}(\boldsymbol{r},0)=0$, then raising the $x-y$ gaussian barrier $V_{\rm barr}(x,y,t)$ up to a height $V_{\rm b}$. The latter is centered at $(x=R_0,y=0)$ with  $1/e^2$ radius $\sigma_x=7.55$ and $\sigma_y=2.15$. 

The initial state at $t=0$ is obtained by finding the ground state of the GP equation by imaginary time propagation with $\ell=0$ and without barrier, and then imprinting one quantum of circulation on the wave function by multiplying by $\exp(i\phi)$ with $\phi$ being the azimuthal coordinate. Then the barrier is linearly ramped up reaching $V_{\rm b}$ at $t=40$ ($\sim 60 {\rm ms}$). The system is then typically evolved up to $t=60$. 

We numerically integrate the 3D GP equation using a finite-difference real space product formula (RSPF) approach
described in \cite{collins_num}.


\section{Critical barrier height}

\begin{figure}
\hspace{2.7cm}
\includegraphics[scale=0.28]{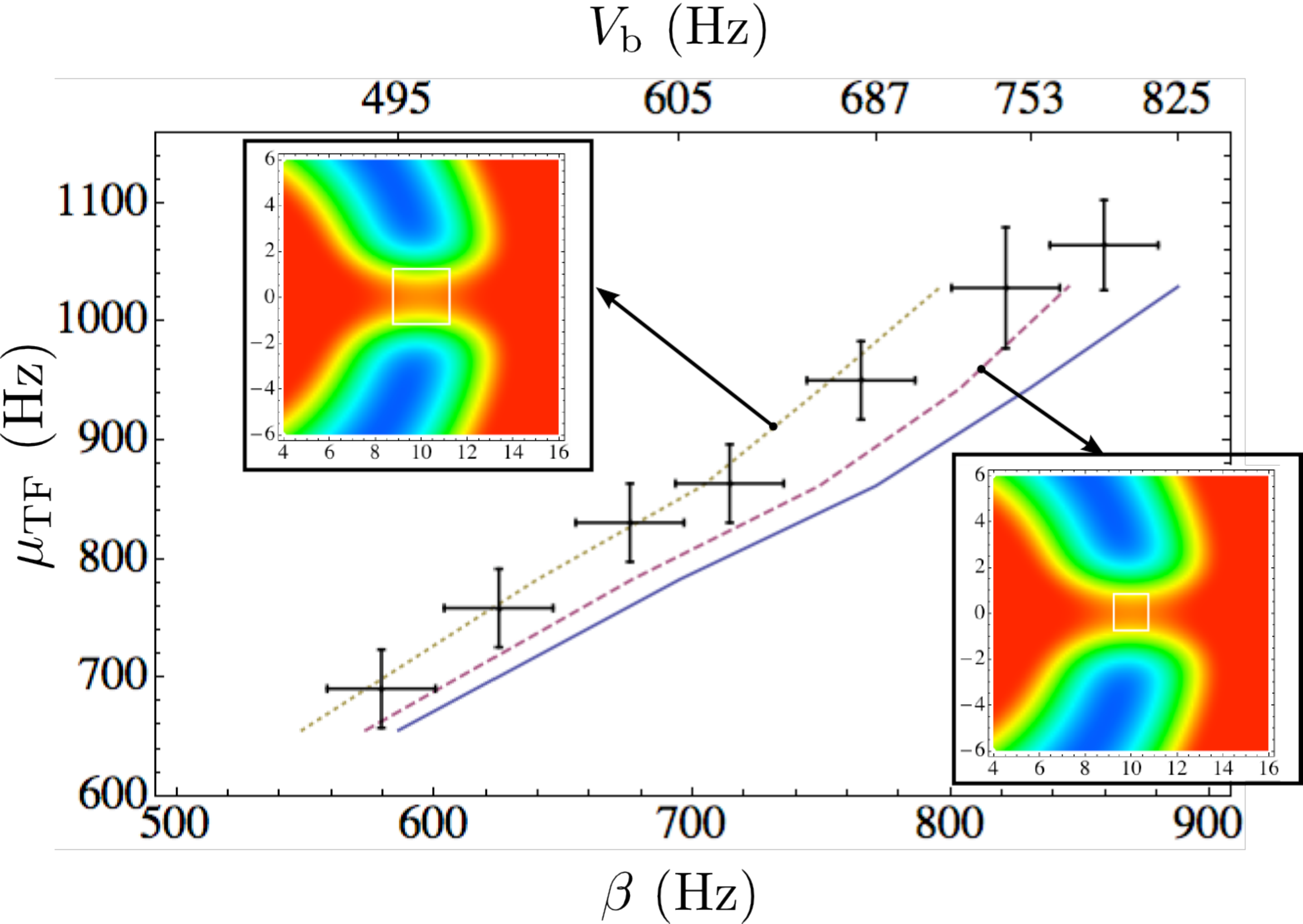}
\caption{(Color online) Chemical potential at the critical point for superfluid instability of the $\ell=1$ state as a function of the chemical potential decrease $\beta$ inside the barrier region. On the upper $x$-axis we report the value of the barrier height $V_{\rm b}$. Black dots with error bars: NIST measurement. Blue solid line: GP prediction. Dashed and dotted lines: GP prediciton corrected for finite imaging resolution using a square averaging region of size $2.5$ and $5$ micrometers for the purple dashed and yellow dotted line, respectively. Insets: column density in the vicinity of the barrier. Squares indicate the averaging region used to produce the curve. $x-y$ coordinates are expressed in harmonic radial trap units.}\label{exp_compare}
\end{figure}

Upon repeating the real time propagation of the GP equation with increasing $V_{\rm b}$, we reach the point where, immediately after the ramping stops, the angular momentum $\ell$ abruptly drops from a value close to one and then stabilizes around zero. The abrupt drop in $\ell$ coincides with a vortex-antivortex pair crossing the annulus and takes place over a very short time (about $0.5 {\rm ms}$). The lowest $V_{\rm b}$ sufficient to observe the drop in $\ell$ defines the critical point.

In Fig.~\ref{exp_compare}, we show the comparison between the GP prediction for the critical point determining the superfluid instability and the experimental measurement performed at NIST. In order to allow for a comparison, we have calculated the chemical potential, both global and local (inside the barrier region) with the same prescription \cite{debug} described in \cite{nist_2011}. 

Given a total atom number $N$, we calculated the global Thomas-Fermi chemical potential $\mu_{\rm TF}$ and subsequently obtained the chemical potential decrease $\beta=\mu_{\rm TF}-\mu_{\rm l}$, where $\mu_{\rm  l}$ is the local chemical potential inside the barrier region. Instead of taking the actual local chemical potential by using $n(x,y,z)$ directly from our numerical results, we inferred it following the NIST protocol, that is, taking the $z$-integrated column density $\nu(x,y)$ at the point $(R_0,0)$, and assuming either a Gaussian or Thomas-Fermi profile along $z$ \cite{gaussian}, which provides a relation between $\nu(R_0,0)$ and $\mu_{\rm l}$. This procedure produces the blue solid line shown in Fig.~\ref{exp_compare}, which does not very well reproduce the experimental results. 

This discrepancy might indicate that the thermal fluctuations induced by the finite temperature can play an important role in the experiment. This fluctuations trigger the vortex nucleation before the critical velocity predicted from the GP equation is reached.
In \cite{mathey_2012}, the same experimental data were compared with the theoretical prediciton of both the GP equation and the truncated Wigner approximation. 
The model setup was a waveguide with periodic boundary condition, and the barrier taken to be constant along the direction transverse to the flow. The inclusion of thermal fluctuations through the Wigner approach provides a systematic effect which shifts the theoretical predictions closer to the experimental data with respect to the GP results.
With an appropriate temperature compatible with the experiment, the Wigner approach was seen to agree slighlty better with the experimental data than the GP equation. 

An an alternative explanation for the discrepancy between theory and experiment, which remains within the framework of GP at zero temperature, can be found in the finite resolution of the imaging system. 
Indeed, if, instead of the column density $\nu(R_0,0)$, we use its average $\nu$ over a square $x-y$ region $A$ centered about $(R_0,0)$ (see purple dashed and yellow dotted lines), we get a better agreement with the experimental data. This averaging process models the effect of a finite imaging resolution.
We emphasize that the value of $\beta$ strongly depends on the size of $A$, that is, it is strongly affected by any limitation to the resolution of the imaging system. In Fig.~\ref{exp_compare}, a square area whose side length ranges from $2.5$ to $5$ micrometers provides good agreement. 
It must anyway be noted that the experimental data reported in Fig.~\ref{exp_compare} already contain a $15\%$ correction to the local density due to some limitations to the imaging resolution. Thus, the finite area correction to our theoretical data should account for possible further limiting factors, not contained in this $15\%$. 



\section{The instability criterion}
\label{criterion}

\begin{figure}
\hspace{3.8cm}
\includegraphics[scale=0.6]{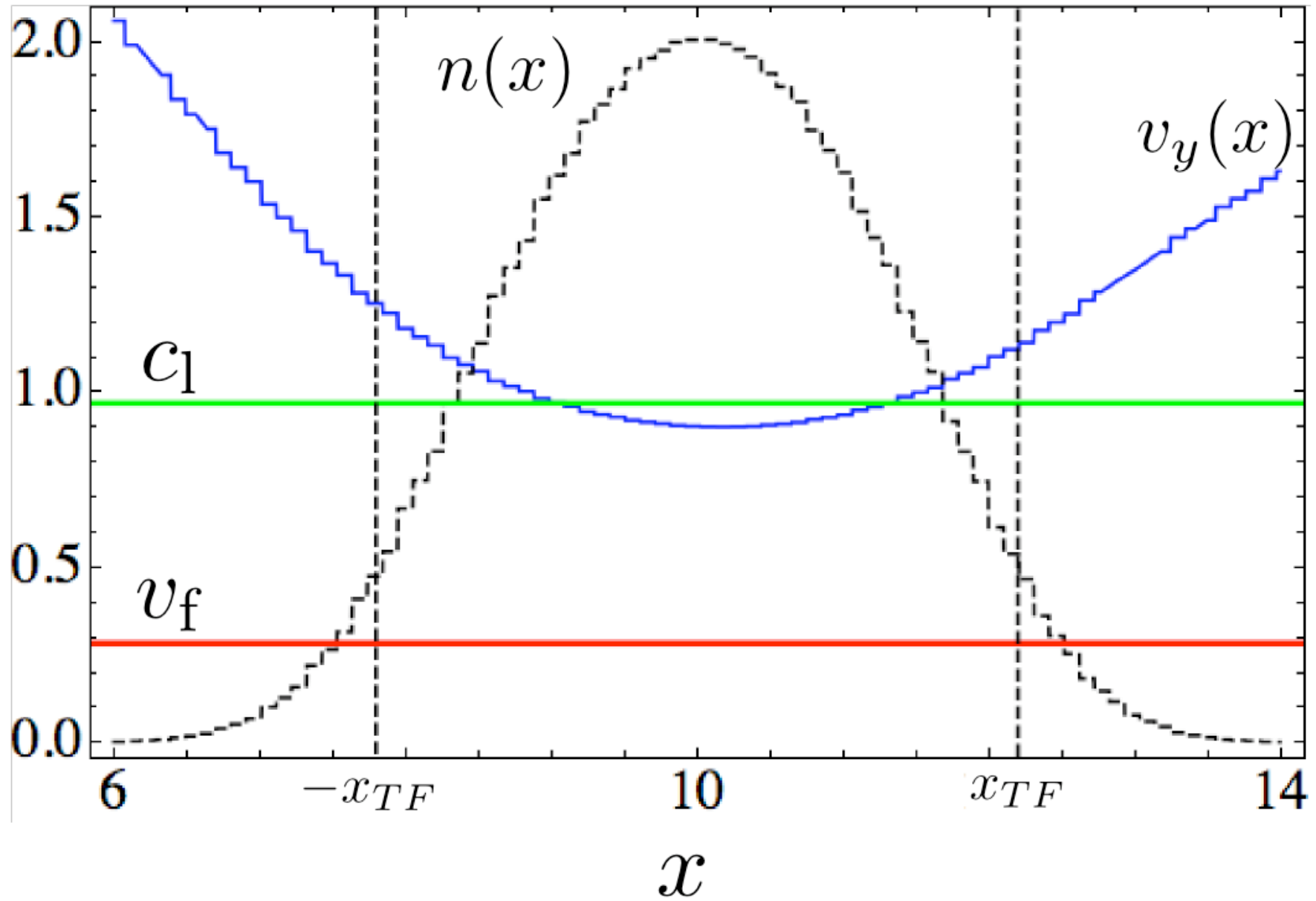}
\caption{(Color online) Fluid velocity at the critical point for superfluid instability of the $\ell=1$ state. Blue solid line: azimuthal component of the local fluid velocity along a radial line crossing the middle of the barrier region. Both the radial coordinate $x$ and the velocity $v(x)$ are expressed in harmonic radial trap units. Black solid line: density (arbitrary units) along the same radial line with vertical dashed black lines indicating the position of the Thomas-Fermi surface. Green solid line: sound velocity inside the barrier region $c_{\rm l}=c_{\rm eff}^{(\rm 2D)}$ for phonons propagating azimuthally in  the $x-y$ plane. Red solid line: Feynman's prediction with $D=5 d_x$ and $a=\xi_{\rm l}$. The chemical potential is $\mu_{TF}=h\times 1030 {\rm Hz}$.} \label{crit_compare}
\end{figure}

We now turn to the discussion of the instability criterion determining the critical point within the GP description.
Putting toghether previous findings involving several different setups and dimensionalities, we start by proposing a general criterion, which should be valid in the hydrodynamic regime of flow.
In this case, the system can be considered locally homogeneous along the flow direction ,that is, the Thomas-Fermi or local density approximation can be employed. We argue that the critical point can be determined using the following criterion: given the excitation spectra calculated in the system without anisotropy along the flow direction, apply the Landau criterion to the anisotropic case locally inside the region where the superfluid velocity is the highest. In the present case, one should take the spectrum in a torus without a barrier \cite{pedri_2012} and calculate the critical velocity, which corresponds to surface mode excitation. Upon treating the barrier in the Thomas-Fermi approximation, one can obtain the spectrum inside the barrier and then apply the Landau criterion. This provides a critical velocity to be compared with the highest fluid velocity, which in this case (see Fig.~\ref{crit_compare}) is reached at the surface of the cloud. 

For instance, it has been verified by simulating the time-dependent GP equation, both in toroidal and in waveguide geometries \cite{piazza_2011}, that the critical point for superfluid instability in the Thomas-Fermi regime is determined by the following condition: the fluid velocity at the Thomas-Fermi surface is approximately equal to the effective sound speed $c_{\rm l}$, that is, the Landau critical velocity for phonon excitation calculated inside the barrier region.
Actually, as metioned above, due to the density inhomogeneity transverse to the flow, the critical velocity for surface modes (excitations with larger momentum localized at the edge of the cloud \cite{anglin_2001}) is lower than the sound speed \cite{shlyapnikov_2001}, and sets therefore the true critical velocity. This also explains why the criterion above involves the superfluid velocity at the Thomas-Fermi surface. Therefore, the use of the sound speed $c_{\rm l}$ is in principle not justified. However, the critical velocity for phonon instability and that for surface-mode instability differ significantly only very deep in the Thomas-Fermi regime \cite{shlyapnikov_2001} with respect to the confinement transverse to the flow (the chemical potential must be larger than ten times the transverse trap frequency). This explains the fact that using the sound speed instead of the critical velocity for surface modes provides a good approximation.

Fig.~\ref{crit_compare} depicts the typical situation for the present setup at the critical point. It shows the velocity profile inside the barrier region (blue solid line) along a radial line ($x=r,\ y=0$) crossing the middle of the barrier. Here the velocity is the largest and therefore the critical condition is first reached. 
The green solid line is the value of the effective sound speed $c_{\rm l}$ corresponding to the above radial line. The effective sound speed, which should be the relevant one in this geometry, is the speed of phonons propagating azimuthally in the $x-y$ plane. It is calculated by considering the degrees of freedom along the $z$ direction as frozen, and then averaging the local 2D sound velocity over the radial direction. We assume a Gaussian profile along $z$, $c_{\rm eff}^{(\rm 2D)}=\sqrt{(2/3)\mu^{(\rm 2D)}/m}$, where $\mu^{(\rm 2D)}=g/(d_z\sqrt{2\pi}) \nu(R_0,0)$, and $\nu(R_0,0)$ is the column density at the center of the barrier.
As we see from Fig.~\ref{crit_compare}, the local criterion described above is not very well verified since, due to the high barrier, the density is strongly depleted at the critical point, and therefore we are outside of the Thomas-Fermi regime. 
Nonetheless, we can see that the local fluid velocity inside the barrier region $v_{\rm barr}$ is still of the order of the effective sound speed $c_{\rm l}$. This is a feature often observed by studying with the GP equation the superfluid instability for a condensate flowing through a constriction (or obstacle) \cite{rica_1992,adams_2000,watanabe_2009,piazza_2011}, provided that the condensate in the bulk region (outside the barrier) be not too far from the Thomas-Fermi regime.

We point out that the increase of the local fluid velocity toward both edges of the annulus inside the barrier is a consequence of the continuity equation: $\nabla(n*\boldsymbol{v})=0$, and is not specific to the torus, but appears already in a waveguide geometry when a barrier is present \cite{piazza_2011}.
The density decreases going toward the edges of the annulus, therefore the velocity has to increase to keep the flux constant.
However, if we had a waveguide without a barrier, that is, simply a linear flow with a non homogeneous density profile transverse to the flow, the stationary solution of the GP equation would have a constant velocity everywhere and no transverse component, because this satisfies the continuity equation anyway. If one instead adds a barrier, in order to satisfy the continuity equation in passing from outside to inside the barrier, the velocity has to increase toward the edges, and also gain a transverse component. In the present case the transverse component is very small. Far away from the barrier, this increase of the fluid velocity toward both edges disappears, and the velocity behaves as in the case without a barrier. In particular, if we are in a toroidal geometry, only the decrease like $1/R$ remains due to the quantization of circulation.

\section{The failure of the Feynman criterion}

The red solid line in Fig.~\ref{crit_compare} corresponds instead to the Feynman prediction for the critical velocity $v_{\rm f}=\hbar/(m D)\ln(D/a)$, where $D$ is the annulus width and $a$ is the vortex core size. For the latter we took the healing length $a=\xi_{\rm l}=(1/\sqrt{2})\hbar/mc_{\rm l}$ corresponding to the effective sound speed $c_{\rm l}$ inside the barrier region. While the width of the annulus is $D\simeq 8 d_x$ outside the barrier region, it is reduced inside: $D\simeq 5 d_x$ (we took the Thomas-Fermi radial width of the torus inside the barrier region). We see that the Feynman prediction $v_{\rm f}$ is significantly lower than fluid velocity inside the barrier region and is even lower than the fluid velocity outside the barrier region $v_{\rm bulk}\simeq 0.7 d_x\omega_x$. In general, independently of the geometry, and at least for angular momenta per atom $\ell<10$, we always found with GP that the Feynman prediction is typically much lower than the fluid velocity in the barrier region. 
The generality of this fact becomes apparent if we rewrite the Feynman velocity as 
\begin{equation}
\label{feyn}
v_{\rm f}= c_{\rm l}\ \sqrt{2}\ \frac{\xi_{\rm l}}{D}\ln\left(\frac{D}{\xi_{\rm l}}\right) =\epsilon\ c_{\rm l} \ . 
\end{equation}
As stated before, the superfluid instability, within the GP equation, corresponds typically to a situation where the local fluid velocity inside the barrier region $v_{\rm barr}$ is of the order of the effective sound speed $c_{\rm l}$. Now, since the expression for the Feynman velocity given above is only valid when the channel width (in our case the annulus width) is much larger than the vortex core size, i.e. $\tilde{D}=D/\xi_{\rm l}\gg 1$, we have $v_{\rm f}=\epsilon\ c_{\rm l}$, where $\epsilon=\sqrt{2}\ln(\tilde{D})/\tilde{D}\ll 1$ is a small parameter. 

Eq.~(\ref{feyn}) indicates that the Feynman prediction is not simply a very rough estimate of the true GP critical velocity but is instead intrinsically different, since it cannot be applied to instabilities whose critical point corresponds to a superfluid velocity being of the order of the sound speed.

On the other hand, the measurements performed at NIST \cite{nist_2011} seemed to suggest that the Feynman prediction may apply. 
As discussed above, this is not compatible with the GP equation, and, if we still want to interpret the experiment with the latter, 
we have to assume that, due to the limitations to the imaging resolution, the density inside the barrier is overestimated from the experimental procedure (see discussion above regarding the comparison of the critical barrier height). This in turn implies an overestimation of the local chemical potential, which in turn overestimates $v_{\rm f}$. One can understand this by using Eq.~(\ref{feyn}) and assuming the Thomas-Fermi description to hold, even though this is not a very good approximation as discussed above. A larger chemical potential implies a smaller healing length and, if we approximately neglect the change in the effective annulus width, an increased $v_{\rm f}$\cite{overestimation}. Secondly, the local fluid velocity inside the barrier is underestimated by solving the equivalent 1D flow problem obtained by integrating radially the measured column density, as done in \cite{nist_2011}. 

 
\section{Phase slip dynamics}

\begin{figure}
\vspace{0.5cm}
\hspace{2.8cm}
\includegraphics[scale=0.6]{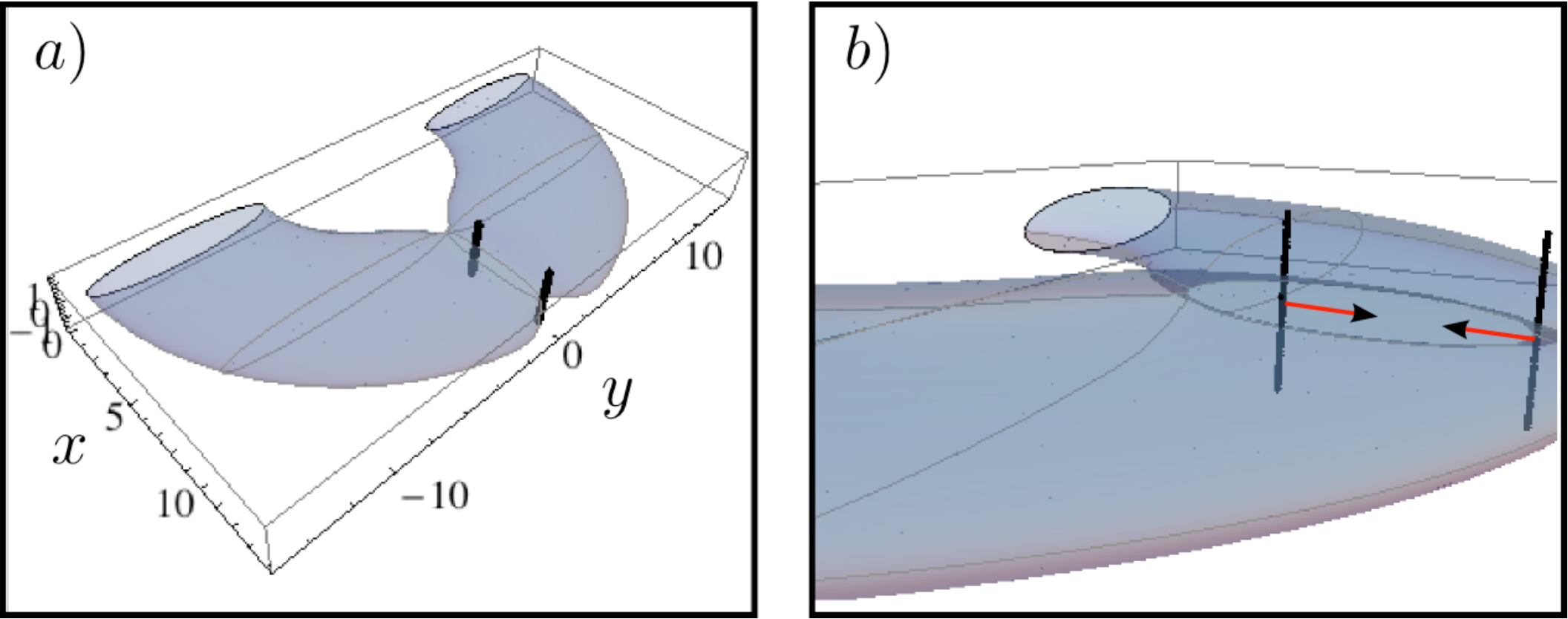}
\caption{(Color online) Superfluid decay dynamics of the $\ell=1$ state. The parameters are the same as in Fig.~\ref{crit_compare}. Black dots indicate the positions of the vortex cores. The gray surface corresponds to the Thomas-Fermi surface of the cloud. a) and b) represent two different viewpoints of the same moment of the dynamics. Two straight vortex lines move toward each other, as indicated by the arrows.}\label{annihilation}
\end{figure}

In Fig.~\ref{annihilation} we show an example of the superfluid decay dynamics resulting from the numerical solution of the GP equation in real time. The positions of the vortex cores, determined using a plaquette method \cite{piazza_2011}, are depicted as black dots while the gray surface represents the Thomas-Fermi surface of the cloud.
Two different views, referring to the same instant after the barrier has reached the critical height, show two straight vortex lines with opposite winding numbers moving toward the center of the annulus. The two lines eventually meet close to the outer edge of the annulus and annihilate, thereby producing a full phase slip that brings the angular momentum to zero. One line comes from the low density region at the torus center while the other moves inwards from the system boundary. As can be observed also from the drop in the angular momentum, the phase slip event is very fast, taking about $0.5 {\rm ms}$ for the lines to cross the annulus and annihilate. The inner line enters the bulk first since the instability is reached first at the inner edge of the annulus due to the velocity asymmetry. The latter is caused by the radial decrease of the velocity due to the quantisation of circulation. In \cite{piazza_2009}, we have shown that this asymmetry can give rise to two separate critical barrier heights, so that below the highest of the two only a vortex enters the annulus from the inner edge. However, in the present case, due to the low angular momentum, the difference between the two critical heights is so small that, in order to resolve these two critical points, we would need to fine tune our barrier height. However, even if we do not resolve the two barrier heights, we see that the inner line enters much before the line coming from the outer edge so that they meet close to the outer edge of the annulus. If we would further decrease the barrier height, we can imagine the point at which the two vortices annihilate to move further and further toward the outer region so that at some point, the annihilation would take place far enough outside the cloud that the outer vortex would not play any role.
This asymmetry characteristic of a torus has been observed to influence the Landau criterion for a BEC in a torus without barrier \cite{pedri_2012}, where the surface modes become unstable first at the inner edge.

Since the decay dynamics involve straight vortex lines, it is reasonable to ask ourselves what prediction we would get if we modify the above Feynman velocity, related to vortex rings, to the case of nucleation of vortex lines. As discussed in \cite{anderson_1966}, the Feynman critical velocity for the nucleation of a vortex line crossing the channel is smaller by a prefactor $2/\pi$ than the critical velocity for ring nucleation, which brings the Feynman prediction even further away from the GP results.


\section{Conclusions}

Our numerical simulations have shown a dicrepancy between the critical barrier measured in \cite{nist_2011} and the theoretical prediction of the zero temperature GP equation. This might be due to the finite temperature which affects the phase slippage dynamics by allowing for the thermal nucleation of vortices. 
As an alternative explanation of the discrepancy we considered the finite resolution of the imaging system, since the latter strongly influences the comparison between theory and experiment.

We have then turned our attention to the instability criterion determining the critical velocity.
Considering the results of our analysis and of the one made in the previous literature, some questions arise: 1) why is the Feynman criterion incompatible with the GP, and what is its relation, if any exists, with the local criterion for the instability of surface modes? 2) What is the nature of the GP instability? Is it an energetic instability, a dynamical one, or something else? 
In the conclusion of this work, we will address these questions and argue that while well posed no clear answers have yet emerged despite their fundamental relevance for the understanding of superfluidity.

1) The physical reason underlying the the failure of the Feynman prediction, observed here with the GP equation and experimentally in \cite{hadzibabic_2012}, is not understood. It can be argued that the Feynman critical velocity is actually only an order of magnitude estimate and its failure does not mean much. However, we showed that the incompatibility is more fundamental (see Eq.~(\ref{feyn})) and that the Feynman criterion cannot provide even the correct order of magnitude. This is true every time the critical velocity is of the order of the sound speed, like it is for the GP equation. On the other hand, one might suppose that the Feynman criterion could work well when additional defects are present inside the barrier region.
In general, a useful insight might be provided by relating the Feynman criterion, based on vortices, to the instability of surface modes \cite{feyn_note}. 


2) The first observation is that we have a quantitatively viable criterion for predicting the critical velocity, as we discussed in section \ref{criterion}. 
This criterion has also been previously theoretically verified to a reasonable degree for the case of a BEC flowing past an obstacle in \cite{rica_1992,adams_2000,watanabe_2009,piazza_2011}, and for a rotating elliptical trap \cite{schneider_1999}, and also for a fermionic superfluid using the Bogoliubov-de Gennes equations \cite{spuntarelli}. Remarkably, it has been even experimentally verified at the Cavendish Laboratory \cite{hadzibabic_2012} where the instability is attributed to the roughness of the trap creating a region where the flow is constricted.
As described above, in general, this criterion is based on the excitation spectra calculated in the system without anisotropy along the flow direction, for instance in a transversally inhomogeneous waveguide \cite{zaremba_1998,stringari_1998,anglin_2001,shlyapnikov_2001}, a rotationally symmetric trap \cite{dalfovo_1997,lundh_1997,machida_1999,fetter_2001}, or a torus without barrier \cite{pedri_2012}. Given the spectrum, one applies the Landau criterion to the anisotropic case locally inside the region where the superfluid velocity is the highest, assuming the local density approximation to hold along the flow direction. 
Therefore, since the fluid is locally homogeneous, it has been suggested in the literature
that the instability is energetic.

However, the second observation is that time-dependent GP simulations give indications that the instability might be dynamical, since:
i) no defects \cite{noise} exist inside the barrier region or in general the ``high velocity region'' which would be required in order to trigger the energetic instability \cite{defects}; and ii) the nucleation dynamics is generally fast, which means that excitations grow very quickly as for typical dynamical instabilities, whose onset can indeed be observed in absence of defects. 

We note that this kind of apparent ambiguity appears also in the studies of the critical angular velocity for a BEC in a rotating harmonic trap. In particular, using a linear stability analysis in the presence of anisotropy, the surface modes have been shown to become dynamically unstable \cite{castin_2001}, and the critical angular frequency to agree with the real-time propagation of the GP equation \cite{martin_2006}. At the same time however, the real-time GP results in the anisotropic trap have been observed \cite{schneider_1999} to be also well predicted by the ``local Landau criterion'' described above. To our knowledge, a conclusive statement about the relation between the local Landau criterion and dynamical instability is still missing in this context.

A particular setup which could help understanding this issue is the one-dimensional condensate flow crossing the speed of sound twice, thereby forming two sonic horizons \cite{finazzi_2010}. In this configuration, a linear stability analysis predicts the existence of a global dynamical instability when the size of the supersonic region becomes larger than a critical value. This hints to a possible link with the Landau criterion applied inside the supersonic region.
The correspondence between the local Landau criterion and the global dynamical instability seems crucial to understand the mechanism underlying vortex nucleation and deserves anyway further study.

\section{Acknowledgments}
We would like to thank A. Recati for fruitful discussions.
F. P. acknowledges support from the Alexander Von Humboldt foundation. A.S. acknowledges support from the EU-STREP Project QIBEC.
This work was supported the Los Alamos National Laboratory, operated by Los Alamos National Security, LLC for the National Nuclear Security Administration of the U.S. Department of Energy under Contract No.~DE-AC52-06NA25396, with computer resources provide  by a LANL Institutional Computing grant.


\section*{References}

\begin{thebibliography}{40}
\bibitem{nist_2007}{C. Ryu, \emph{et al.}, Phys. Rev. Lett {\bf 99}, 260401(2007)}
\bibitem{nist_2011}{A. Ramanathan, \emph{et al.}, Phys. Rev. Lett {\bf 106}, 130401 (2011)}
\bibitem{hadzibabic_2012}{S. Moulder, \emph{et al.}, Phys. Rev. A {\bf 86}, 013629 (2012)}
\bibitem{leggett_rmp_1999}{A. J. Leggett, Rev. Mod. Phys. {\bf 71}, 318 (1999)}
\bibitem{clarke_2004}{J. Clarke, and A. Braginski,\emph{The SQUID Handbook} (Wiley-VCH, Weinheim, 2004), Vols. 1,2}
\bibitem{packard_2012}{Y. Sato, and R. E. Packard, Rep. Prog. Phys. {\bf 75}, 016401 (2010)}
\bibitem{nist_2012}{K. C. Wright, \emph{et al.}, arXiv:1208.3608v1}
\bibitem{ketterle_1999}{C. Raman, \emph{et al.}, Phys. Rev. Lett {\bf 83}, 2502 (1999)}
\bibitem{ketterle_2000}{R. Onofrio, \emph{et al.}, Phys. Rev. Lett {\bf 85}, 2228 (2000)}
\bibitem{engels_2007}{P. Engels, and C. Atherton, Phys. Rev. Lett {\bf 99}, 160405 (2007)}
\bibitem{varoquaux_2006}{E. Varoquaux, C. R. Phys. {\bf 7}, 1101 (2006)}
\bibitem{piazza_2009}{F. Piazza, L. A. Collins, and A. Smerzi, Phys. Rev. A {\bf 80}, 021601 (2009)}
\bibitem{mathey_2012}{A. C. Mathey, C. W. Clark, L. Mathey, arXiv:1207.0501 (2012)}
\bibitem{rica_1992}{T. Frisch, Y. Pomeau, and S. Rica, Phys. Rev. Lett {\bf 69}, 1644 (1992)}
\bibitem{adams_2000}{T. Winiecki, B. Jackson, J. F. McCann, and C. S. Adams, J. Phys. B {\bf 33}, 4069 (2000)}
\bibitem{watanabe_2009}{G. Watanabe \emph{et al.}, Phys. Rev. A {\bf 80}, 053602 (2009)}
\bibitem{piazza_2011}{F. Piazza, L. A. Collins, and A. Smerzi, New J. Phys. {\bf 13}, 043008 (2011)}
\bibitem{feynman_1955}{R. Feynman, Prog. Low Temp. Phys. {\bf 1}, 17 (1955)}.
\bibitem{anglin_2001}{J. R. Anglin, Phys. Rev. Lett. {\bf 87}, 240401 (2001)}.
\bibitem{collins_num}{B. I. Schneider, L. A. Collins, and S. X. Hu, Phys. Rev. E {\bf 73}, 036708 (2006)}
\bibitem{debug}{We corrected for a missing factor $1/\sqrt{2}$ in \cite{nist_2011}, see also \cite{mathey_2012}}
\bibitem{gaussian}{In all the cases explored, we used a Gaussian profile since $\mu_{\rm l}$ was always smaller than $\omega_z$}
\bibitem{pedri_2012}{R. Dubessy, \emph{et al.}, Phys. Rev. A {\bf 86}, 011602(R) (2012)}.
\bibitem{shlyapnikov_2001}{P. O. Fedichev, and G. V. Shlyapnikov, Phys. Rev. A {\bf 63}, 045601 (2001)}
\bibitem{overestimation}{Since both the healing length and the effective annulus width are modified by a change in the chemical potential, it is not true in general that an increase in the chemical potential makes the Feynman velocity increase, but rather holds only for sufficiently small chemical potentials, as is the case here.}
\bibitem{anderson_1966}{P. W. Anderson, Rev. Mod. Phys. {\bf 38}, 298 (1966)}
\bibitem{feyn_note}{For a homogeneous condensate flowing past a very large and impenetrable obstacle, a linear stability analysis provided a critical velocity in agreement with the Feynman criterion \cite{zwerger_2000}. However, time-dependent simulations in this setup \cite{rica_1992, adams_2000} have verified that the instability corresponds instead to the local fluid velocity reaching the sound speed. This might be explained by the fact that the linear stability analysis of \cite{zwerger_2000} signals an energetic instability which is however not triggered in the time-dependent simulations of \cite{rica_1992, adams_2000}, due to the absence of defects.}
\bibitem{zwerger_2000}{ J. S. Stiessberger, and W. Zwerger, Phys. Rev. A {\bf 62}, 061601(R) (2000)}
\bibitem{schneider_1999}{D. L. Feder, C. W. Clark, and B. I. Schneider, Phys. Rev. A {\bf 61}, 011601(R) (1999)}
\bibitem{spuntarelli}{A. Spuntarelli, P. Pieri, and G. C. Strinati, Physics Reports {\bf 488}, 111 (2010)}
\bibitem{zaremba_1998}{E. Zaremba, Phys. Rev. A {\bf 57}, 518 (1998)}
\bibitem{stringari_1998}{S. Stringari, Phys. Rev. A {\bf 58}, 2385 (1998)}
\bibitem{dalfovo_1997}{F. Dalfovo, \emph{et al.}, Phys. Rev. A {\bf 56}, 3840 (1997)}
\bibitem{lundh_1997}{E. Lundh, C. J. Pethick, H. Smith, \emph{et al.}, Phys. Rev. A {\bf 55}, 2126 (1997)}
\bibitem{machida_1999}{T. Isoshima, K. Machida, Phys. Rev. A {\bf 60}, 3313 (1999)}
\bibitem{fetter_2001}{D. L. Feder, \emph{et al.}, Phys. Rev. Lett. {\bf 86}, 564 (2001)}.
\bibitem{noise}{The finite resolution in the computation generates numerical noise which however should not be sufficient to trigger an energetical instability within our computational times.} 
\bibitem{defects}{Note that, within the local Landau criterion, the macroscopic obstacle `''disappears'' as its effect is included as a modification of the parameters of the locally homogeneous system. Therefore, we need additional defects inside the obstacle region in order to trigger the energetic instability in a non perturbed homogeneous system. See A. Leggett, Rev. Mod. Phys. {\bf 73}, 307 (2001), I. Carusotto, \emph{et al.}, Phys. Rev. Lett. {\bf 97}, 260403 (2006)}
\bibitem{castin_2001}{S. Sinha, and Y. Castin, Phys. Rev. Lett {\bf 87}, 190402 (2001)}
\bibitem{martin_2006}{N. G. Parker, R. M. W. van Bijnen, and A. M. Martin, Phys. Rev. A {\bf 73}, 061603(R) (2006)}
\bibitem{finazzi_2010}{S. Finazzi, and R. Parentani, New J. Phys. {\bf 12}, 095015 (2010)}

\end{thebibliography}
\end{document}